\documentclass[12pt]{article}
\usepackage{epsf}
\usepackage{amsmath}
\usepackage{graphics}
\usepackage{cite}

\renewcommand{\thefootnote}{\#\arabic{footnote}}
\begin{document}

\newcommand{\gtrsim}{ \mathop{}_{\textstyle \sim}^{\textstyle >} }
\newcommand{\lesssim}{ \mathop{}_{\textstyle \sim}^{\textstyle <} }

\newcommand{\rem}[1]{{\bf #1}}

\renewcommand{\thefootnote}{\fnsymbol{footnote}}
\setcounter{footnote}{0}
\begin{titlepage}

\def\thefootnote{\fnsymbol{footnote}}

\begin{center}

\vskip .5in

\bigskip
\bigskip
{\Large \bf Prediction for $\Gamma(H\rightarrow \tau^+\tau^-)/
\Gamma(H\rightarrow \mu^+\mu^-)$ from Non-Abelian 
Flavor Symmetry}

\vskip .45in

{\bf Paul H. Frampton\footnote{frampton@physics.unc.edu}
and Shinya Matsuzaki\footnote{synya@physics.unc.edu}}

\bigskip
\bigskip

{Department of Physics and Astronomy, UNC-Chapel Hill, NC 27599, USA.}

\end{center}

\vskip .4in

\begin{abstract}
It is shown that the ratio of branching ratios in Higgs boson decay
$r = \Gamma(H \rightarrow \tau^+\tau^-)/\Gamma(H\rightarrow \mu^+\mu^-)$
in a simplified ($T^{'} \times Z_2$) model is $(r)_{T^{'} \times Z_2}
\simeq 0.001$.
This prediction is on a footing with the successful one in 
$(s, d)$ mixing, $\tan 2 \Theta_{12} = \frac{1}{3}(\sqrt{2})$.
This value 
for $r$ differs from that of the minimal standard model
by a very large factor (over 250,000)
and can provide a smoking gun for ($T^{'} \times Z_2$) flavor symmetry.
\end{abstract}

\end{titlepage}

\renewcommand{\thepage}{\arabic{page}}
\setcounter{page}{1}
\renewcommand{\thefootnote}{\#\arabic{footnote}}

\newpage

\bigskip

Unification is a common thread in theoretical physics.
One example is field theory which is used in an acceleratedly expanding
number of areas ranging
from
particle physics where it had its first
application to nuclear physics, cosmology and condensed matter physics
among others.

\bigskip

In particle physics, the minimal standard model (MSM)
is a field theory
for quarks and leptons which is
phenomenologically very successful yet has two aspects which display
lack of a what might be reasonably expected unification.

\bigskip

The more obvious of the two is that there are independent
gauge coupling constants corresponding to the factor groups
$SU(3) \times SU(2) \times U(1)$ which may be labeled as
$g_3, g_2, g_1$.
Although electromagnetic and weak interactions
are partially unified in the
$SU(2) \times U(1)$ electroweak theory, there are two
couplings $g_2, g_1$ which
are unrelated until
the electroweak mixing angle $\theta_{EW}$ is specified.
The value of $\theta_{EW}$ is empirical input.

\bigskip

This lack of unification of coupling constants
was addressed by grand unified theories (GUTs).
The simplest GUT is based on $SU(5)$ \cite{GG}
and envisages a GUT energy scale orders of magnitude
higher than the electroweak scale ($M_{GUT} \simeq 10^{16}$ GeV
compared to $M_{EW} \simeq 10^2$ GeV) and a "desert"
without new physics between these two scales.
The first support for GUT came from its capability
to suggest relationships between quarks and leptons.
For example\cite{BEGN} it was pointed out that
the GUT scale prediction $m_b = m_{\tau}$ is renormalized
to $m_b \simeq 3 m_{\tau}$ at low energy, in reasonable
agreement with experiment.

\bigskip

An $SU(5)$ GUT made other important predictions
including for $\theta_{EW}$ and the proton decay
lifetime $\tau_p$. The value $\sin \theta_{EW} = 3/8$
at the GUT scale is renormalized at low energy to a value
unacceptably smaller than the measured value.
Proton decay has not been seen and the lower bound
on $\tau_p$ is some orders of magnitude higher than the
$SU(5)$ prediction. The acceptable value for $m_b/m_{\tau}$
must regrettable now be regarded as fortuitous because 
the other predictions fail.
The $SU(5)$ predictions for $\theta_{EW}$ and $\tau_p$ can
be corrected by {\it ad hoc} modification\cite{FG} of the
theory but this accommodates arbitrary values rather
than leading to a predictive theory.

\bigskip

Grand unification received a new lease of life with the introduction
of supersymmetry (Susy). The generalization of GUTs to supersymmetry
(SusyGUTs) can accommodate both acceptable $\theta_{EW}$ and $\tau_p$;
so low energy SUSY may show up
at the L.H.C.

\bigskip

The reasons for optimism about discovery of
supersymmetry at the TeV scale are balanced by the fact
that the three key advantages 
of low-energy SUSY -- (i) cancellation of quadratic divergences
in the MSM (its original motivation); (ii) improved unification
in SusyGUTs compared to GUTs; (iii) an attractive dark matter candidate,
the neutralino as WIMP -- can equally be achieved without SUSY,
for example, using conformality\cite{review} which
leads to different solutions of the same issues.

\bigskip

The second of the two lacks of unification in the standard
model is a concern already alluded to, the unity of
leptons and quarks. Except for the quark-lepton correspondence 
in the three families in that
there are doublets
$(t, b), (c, s), (u, d)$ quarks and
$(\nu_{\tau}, \tau), (\nu_{\mu}, \mu), (\nu_e, e)$ leptons,
there is no established relationship
between quarks and leptons. This is a striking fact.

\bigskip

Therefore we study an alternative to grand unification.
Instead of the attempt to unify couplings by {\it e.g.}

\begin{equation}
(SU(3) \times SU(2) \times U(1)) \subset G_{GUT} 
\label{GUT}
\end{equation}
we entertain a flavor symmetry

\begin{equation}
(SU(3) \times SU(2) \times U(1)) ~ {\rm and} ~  G_{Flavor}
\label{FlavorSymmetry}
\end{equation}
where $G_{Flavor}$ is spontaneously broken at a TeV scale.
The idea is that behavior 
under $G_{Flavor}$ will lead
from the group theory to novel relationships
between quarks and leptons.

\bigskip

What is the best choice for $G_{Flavor}$? Generally
an infinite Lie group for $G_{Flavor}$ would
be expected to be gauged (otherwise it will
not be respected by gravity) and this leads to
additional gauge bosons effectively
extending the MSM gauge group in
Eq.(\ref{FlavorSymmetry}). But with $G_{Flavor}$ necessarily
not commuting with the other factors because
particles which have the same 3-2-1 quantum numbers
will transform differently under $G_{Flavor}$
this cannot be of the GUT form in Eq.(\ref{GUT}). The
peculiar way of writing ``and" in Eq.(\ref{FlavorSymmetry}) reflects
that $G_{Flavor}$ will be an overarching broken
global symmetry aimed only to relate the 
parameters in the MSM.

\bigskip

Finite groups are either Abelian or non-Abelian. The Abelian varieties
have irreps which are all one-dimensioal and therefore insufficiently
structured for fruitful model building.

\bigskip

On the other hand, {\it all} non-Abelian finite groups (with
doublet, triplet, etc. irreps)
have been presented up to
order $g \leq 31$ in \cite{FKgroups}. There are exactly
45 such groups. The choice
of $G_{Flavor}$ can be narrowed
by the fact that
experimental data\cite{PDG2008} on the neutrino
mixing angles give values $\theta_{12} \simeq 34^o$, $\theta_{23}
\simeq 45^o$ and $\theta_{13} < 13^o$ leading\cite{HPS} to the
tribimaximal mixing (TBM) ansatz $\tan \theta_{12} = 1/\sqrt{2}$,
$\theta_{23} = 45^o$ and $\theta_{13} = 0$. Such
a TBM may be underwritten\cite{Ma} by a leptonic flavor symmetry
$G_{Flavor} = A_4$.

\bigskip

Extension of $A_4$ itself to quarks in Eq.(\ref{FlavorSymmetry})
is problematic. The basic reason is that the neutrinos
have Majorana masses while quarks have Dirac masses.
Nevertheless, there is available the double cover $T^{'}$
of $A_4$ which is a subgroup of $SU(2)$: $T^{'} \subset SU(2)$
as $A_4$ is a subgroup of $SO(3)$: $A_4 \subset SO(3)$.
This leads to the adoption of $G_{Flavor} = (T^{'} \times Z_2)$
as in \cite{Tprime} where the extra $Z_2$ is necessary, just
like R-symmetry in Susy models, to exclude phenomenologically
unacceptable Yukawa couplings.

\bigskip

The use of ($T^{'} \times Z_2$) already led to one successful
prediction for the Cabibbo angle, in \cite{Tprime} where 
the calculations were made more tractable in
a simplified model where the $b$ quark is stable.
This is a reasonable approximation to reality because
the quark mixings satisfy $\Theta_{12} \gg \Theta_{23} \gg \Theta_{13}$.
Thus the corrections, as confirmed by further calculation
\footnote{work in progress.}
are small. he prediction of $\Theta_{12}$ arises because of a
messenger scalar linking the three
neutrinos $(\nu_{\tau}, \nu_{\mu}, \nu_e)_L$ to
the D-type quarks of the first two generations
which are accommodated in $Q_L (2_1, +1) ~ [(c, s)_L, (u, d)_L]$ and
${\cal S}_R (2_2, +1) ~ [(s, d)_R]$. 

\bigskip

\noindent Before embellishing this simplified model, we here
draw attention to a striking prediction it contains
which, if confirmed by experiment, would provide encouragement
to this direction.

\bigskip

We recall the general features of the model.
The leptons are assigned under ($T^{'} \times Z_2$) as
\begin{equation}
\begin{array}{ccc}
\left. \begin{array}{c}
\left( \begin{array}{c} \nu_{\tau} \\ \tau^- \end{array} \right)_{L} \\
\left( \begin{array}{c} \nu_{\mu} \\ \mu^- \end{array} \right)_{L} \\
\left( \begin{array}{c} \nu_e \\ e^- \end{array} \right)_{L}
\end{array} \right\}
L_L  (3, +1)  &
\begin{array}{c}
~ \tau^-_{R}~ (1_1, -1)   \\
~ \mu^-_{R} ~ (1_2, -1) \\
~ e^-_{R} ~ (1_3, -1)  \end{array}
&
\begin{array}{c}
~ N^{(1)}_{R} ~ (1_1, +1) \\
~ N^{(2)}_R ~ (1_2, +1) \\
~ N^{(3)}_{R} ~ (1_3, +1),\\  \end{array}
\end{array}
\end{equation}
Imposing renormalizability on the lepton lagrangian
allows as nontrivial terms
only Majorana mass terms and Yukawa couplings
to $T^{'}$ Higgs scalars
\footnote{All scalars are doublets under electroweak $SU(2)$.}
$H_3(3,+1)$ and $H_3^{'}(3,-1)$ 

\begin{eqnarray}
{\cal L}_Y^{(leptons)}
&=&
\frac{1}{2} M_1 N_R^{(1)} N_R^{(1)} + M_{23} N_R^{(2)} N_R^{(3)} \nonumber \\
& & + \Bigg\{
Y_{1} \left( L_L N_R^{(1)} H_3 \right) + Y_{2} \left(  L_L N_R^{(2)}  H_3
\right) + Y_{3}
\left( L_L N_R^{(3)} H_3 \right)  \nonumber \\
&& +
Y_\tau \left( L_L \tau_R H'_3 \right) 
+ Y_\mu  \left( L_L \mu_R  H'_3 \right) +
Y_e \left( L_L e_R H'_3 \right) 
\Bigg\}
+
{\rm h.c.}.
\label{Ylepton}
\end{eqnarray}

\bigskip

\noindent Charged lepton masses arise from the vacuum expectation value
(hereafter VEV)

\begin{equation}
<H_3^{'}> =
\left(\frac{m_{\tau}}{Y_{\tau}},\frac{m_{\mu}}{Y_{\mu}},\frac{m_{e}}{Y_{e}}
\right) = ( M_{\tau}, M_{\mu}, M_e ),
\label{Hprime}
\end{equation}
where $M_i \equiv m_i/Y_i$ ($i = \tau, \mu, e$).

\bigskip

\noindent Neutrino masses and mixings 
come from the see-saw mechanism
and the VEV

\bigskip

\begin{equation}
<H_3> = V( 1, -2, 1).
\label{VEV}
\end{equation}

\newpage

\noindent Left-handed quark doublets \noindent $(t, b)_L, (c, d)_L, (u, d)_L$
are assigned under $(T^{'} \times Z_2)$ to

\begin{equation}
\begin{array}{cc}
\left( \begin{array}{c} t \\ b \end{array} \right)_{L}
~ {\cal Q}_L ~~~~~~~~~~~ ({\bf 1_1}, +1)   \\
\left. \begin{array}{c} \left( \begin{array}{c} c \\ s \end{array} \right)_{L}
\\
\left( \begin{array}{c} u \\ d \end{array} \right)_{L}  \end{array} \right\}
Q_L ~~~~~~~~ ({\bf 2_1}, +1),
\end{array}
\label{qL}
\end{equation}

\noindent and the six right-handed quarks as

\begin{equation}
\begin{array}{c}
t_{R} ~~~~~~~~~~~~~~ ({\bf 1_1}, +1)   \\
b_{R} ~~~~~~~~~~~~~~ ({\bf 1_2}, -1)  \\
\left. \begin{array}{c} c_{R} \\ u_{R} \end{array} \right\}
{\cal C}_R ~~~~~~~~ ({\bf 2_3}, -1)\\
\left. \begin{array}{c} s_{R} \\ d_{R} \end{array} \right\}
{\cal S}_R ~~~~~~~~ ({\bf 2_2}, +1).
\end{array}
\label{qR}
\end{equation}

\noindent Two scalars $H_{1_1} (1_1, +1)$ and
$H_{1_3} (1_3, -1)$ with VEVs
\begin{equation}
<H_{1_1}> = m_t/Y_t, ~~~~ <H_{1_3}> = m_b/Y_b,
\label{H13VEV}
\end{equation}
provide the $(t, b)$ masses.

\bigskip

\noindent The allowed quark Yukawa and mass terms are

\begin{eqnarray}
{\cal L}_Y^{(quarks)}
&=& Y_t ( \{{\cal Q}_L\}_{\bf 1_1}  \{t_R\}_{\bf 1_1} H_{\bf 1_1}) \nonumber \\
&&
+ Y_b (\{{\cal Q}_L\}_{\bf 1_1} \{b_R\}_{\bf 1_2} H_{\bf 1_3} ) \nonumber \\
&& 
+ Y_{{\cal C}} ( \{ Q_L \}_{\bf 2_1} \{ {\cal C}_R \}_{\bf 2_3} H^{'}_{\bf 3})
\nonumber \\
&& 
+ Y_{{\cal S}} ( \{ Q_L \}_{\bf 2_1} \{ {\cal S}_R \}_{\bf 2_2} H_{\bf 3})
\nonumber \\
&&
+ {\rm h.c.}.
\label{Yquark}
\end{eqnarray}

\bigskip

\noindent The precise forms of the couplings in 
Eqs.(\ref{Ylepton},\ref{Yquark}) led to the prediction
for the Cabibbo angle

\begin{equation}
\tan 2\Theta_{12} = \left( \frac{\sqrt{2}}{3} \right)
\label{Cabibbo}
\end{equation}

\bigskip

\noindent The same couplings lead to an even more remarkable prediction
when we study the messenger scalar linking the
charged leptons $(\tau, \mu, e)_L$ to the
U-type quarks contained in $Q_L$ and
${\cal C}_R (2_3, -1) ~ [(c, u)_R]$. This gives rise
to an expression for the ratio of branching ratios
for Higgs decay

\begin{equation}
r = \left( \frac{\Gamma(H \longrightarrow \tau^+\tau^-)}
{\Gamma(H \rightarrow \mu^+\mu^-)} \right)
\label{r}
\end{equation}

\noindent We recall that in the minimal standard model
the two body decays in Eq.(\ref{r}) satisfy at tree level

\begin{equation}
(r)_{SM} = \left( \frac{ m_{\tau}^2 (1 - 4(m_{\tau}^2/M_H^2))^{3/2}}
{m_{\mu}^2 (1 - 4 (m_{\mu}^2/M_H^2))^{3/2}} \right) \simeq 280 
\label{rMSM}
\end{equation}

\noindent In the $(T^{'} \times Z_2)$ model the messenger scalar
$H^{'}_3(3, -1)$ which couples both to neutrinos
and to U-type quarks in the first two generations
provides a large change from Eq.(\ref{rMSM}).
From Eq.(4), we readily calculate the branching ratio in the 
($T^{'} \times Z_2$) model to arrive at

\begin{equation}
\left( \frac{ m_{\tau}^2 (1 - 4(m_{\tau}^2/M_H^2))^{3/2}}
{m_{\mu}^2 (1 - 4 (m_{\mu}^2/M_H^2))^{3/2}} \right)
\left( \frac{M_{\mu}^2}{M_{\tau}^2} \right)
\label{rTprime1}
\end{equation}
where $M_{\tau,\mu}$ are defined in Eq.(\ref{Hprime}), and hence

\begin{equation}
\left| \frac{Y_{\tau}}{Y_{\mu}} \right|^2 = \left( \frac{m_{\tau}^2}{m_{\mu}^2} \right)
\left| \frac{M_{\mu}}{M_{\tau}} \right|^2
\label{YY}
\end{equation}

\noindent In the expression Eq.(\ref{rTprime1}) the dependence on the mixing angles,
which can arise from diagonalizing the Higgs sector,
cancels between the numerator and denominator.
It is crucial to note that the messenger field $H_3^{'}(3, -1)$ couples
to the up-type quarks
of the first two generations with the resultant masses

\begin{equation}
|m_u| = \sqrt{\frac{2}{3}} |M_{\mu}|, ~~~~ |m_c| = \sqrt{\frac{2}{3}} |M_{\tau}|
\label{mumc}
\end{equation}
so that we reach

\begin{equation}
(r)_{T^{'}} = 
\left( \frac{ m_{\tau}^2 (1 - 4(m_{\tau}^2/M_H^2))^{3/2}}
{m_{\mu}^2 (1 - 4 (m_{\mu}^2/M_H^2))^{3/2}} \right) 
\left( \frac{m_u^2}{m_c^2} \right)
\simeq 0.001
\label{rTprime}
\end{equation}

\bigskip

\noindent The change in $r$ from the minimal standard model to
$T^{'}$ flavor symmetry is more than a factor 250,000!!!

\bigskip

\noindent The derivation of the result, Eq.(\ref{rTprime}),
follows from the couplings of the scalar $H^{'}_3$
in Eqs.(\ref{Ylepton},\ref{Yquark}) which require
that the relevant component in the $T^{'}$-decompostion
of the Higgs scalar couples
to ($u, \tau$) and ($c, \mu$) masses
respectively. This is predicated by the group
structure of the ($T^{'} \times Z_2$) model with a 
consequent prediction extremely
different from the standard model where the Higgs
coupling is proportional to mass.

\bigskip

Since the Higgs boson and its decay branching ratios
are targets of opportunity for the L.H.C., Eq.(\ref{rTprime})
can provide a smoking gun for such a quark-lepton
relationship arising from $(T^{'} \times Z_2)$ flavor symmetry.

\bigskip

We may ask about the predictions for the separate
numerator and denominator in Eq.(\ref{r}). Further
study reveals that only the ration $r$ is firmly
predicted because the separate decay modes depend
on the precise identification of the light Higgs
doublet among the scalars of the $(T^{'} \times Z_2)$
model. In the ratio $r$, however, this uncertainty
cancels out.

\bigskip

The prediction, Eq.(\ref{rTprime}),
arises from an alternative to grand unification
which is a flavor symmetry with sufficient structure
to relate leptons and quarks. 

\bigskip

It will be amusing to see from experimental data about Higgs boson
decays whether Nature chooses such a broken symmetry
in relating quarks and leptons.

\bigskip

The unity of quarks and leptons was a principal goal
of grand unified theories and led to the prediction
of proton decay as well as expectations for neutrino
masses and the electroweak mixung angle. The central
idea was to subsume the established gauge symmetry
into a larger simple symmetry group
as in Eq.(\ref{GUT}).

\bigskip

The alternative to grand unification, with the
same goal of relating leptons and quarks, is
the use of a global finite Non-Abelian flavor
symmetry such as the $(T^{'} \times Z_2)$ espoused
here.

\bigskip

We have seen that a striking prediction concerns the
leptonic decays of the Higgs boson. In the
standard model the couplings of the Higgs
to fermion pairs goes like the fermion mass and
the two-body decay rates $H \longrightarrow f\bar{f}$
go like $\Gamma(H \longrightarrow f\bar{f})
\propto m_f^2$ up to small phase space corrections.
The Non-Abelian flavor symmetry further contrains
the Higgs coupling and strikingly changes
the ratios of Higgs decay fractions so that
$\tau^+\tau^-$ relative to $\mu^+\mu^-$
is suppressed by a ratio of quark masses
$(m_u/m_c)^2 \simeq 4 \times 10^{-6}$.

\bigskip

If such an effect is observed, and it would seem
difficult to miss once the Higgs boson is discovered,
it will provide strong evidence, a smoking gun,
for this alternative to grand unification.

\bigskip
\bigskip
\bigskip
\bigskip
\bigskip
\bigskip
\bigskip
\bigskip
\bigskip
\bigskip

\begin{center}
{\bf Acknowledgement}
\end{center}

\noindent This work was supported by U.S. Department of Energy grant number
DE-FG02-06ER41418.


\newpage

\bigskip
\bigskip
\bigskip
\bigskip
\bigskip






\begin{thebibliography}{99}
\bibitem{GG}
H.Georgi and S.L. Glashow, Phys. Rev. Lett. {\bf 32,} 438 (1974).
\bibitem{BEGN}
A.J. Buras, J.R. Ellis, M.K. Gaillard and D.V. Nanopoulos,
Nucl. Phys. {\bf B135,} 66 (1978).
\bibitem{FG}
P.H. Frampton and S.L. Glashow, Phys. Lett. {\bf B131,} 340 (1983).
\bibitem{review}
P.H. Frampton and T.W. Kephart, Phys. Reports. {\bf 454,} 203 (2008).
{\tt arXiv:0706.1186 [hep-ph]}.
\bibitem{FKgroups}
P.H. Frampton and T.W. Kephart, Int. J. Mod. Phys. {\bf A10,} 4689 (1995).
{\tt hep-ph/9409330}.
\bibitem{PDG2008}
Particle Data Group, Phys. Lett. {\bf B667,} 1 (2008).
\bibitem{HPS}
P.F. Harrison, D.H. Perkins and W.G. Scott,
Phys. Lett. {\bf B350,} 167 (2002).
{\tt hep-ph/0202074}.
\bibitem{Ma}
E. Ma and G. Rajasekaran, Phys. Rev. {\bf D64,} 113012 (2001).
{\tt hep-ph/0106291}
\bibitem{Tprime}
P.H. Frampton, T.W. Kephart and S. Matsuzaki.
Phys. Rev. {\bf D78,} 073004 (2008). {\tt arXiv:0807.4713 [hep-ph]}.
\end{thebibliography}
\end{document}